\documentclass[preprint,10pt,5p]{elsarticle}
\graphicspath{{figures/}}
\usepackage{hyperref}

\usepackage{atlasphysics} 

\journal{Physics Letters B}

\begin{document}

\begin{frontmatter}
  

\title{Measurement of the production cross section for W-bosons in association
  with jets in $pp$ collisions at $\sqrt{s} = 7$ \TeV~
  with the ATLAS detector}

\author{The ATLAS Collaboration}


\begin{abstract}
  This Letter reports on a first measurement 
of the inclusive $W$+jets cross section 
in proton-proton collisions at a centre-of-mass energy
of 7 \TeV~at the LHC, with the ATLAS detector.
Cross sections, in both the electron and muon decay modes of the
$W$ boson, are presented as a function of jet
multiplicity and of the transverse momentum of the leading and
next-to-leading jets in the event.
Measurements are also
presented of the ratio of 
cross sections
$\sigma(W+ \ge n)/\sigma(W+ \ge n-1)$
for inclusive jet multiplicities $n=1\rm{-}4$.
The results,
based on an integrated luminosity of 1.3 \ipb,
have been corrected for all known detector effects and
are quoted in a limited and well-defined range of jet and lepton
kinematics.
The measured cross sections are compared to particle-level predictions based
on perturbative QCD.  
Next-to-leading order calculations, studied here for $n \le 2$, are
found in good agreement with the data. Leading-order multiparton
event generators, normalized to the NNLO total cross section, describe
the data well for all measured jet multiplicities.

\end{abstract}

\end{frontmatter}

\section{Introduction}

The study of massive vector boson ($V$, where $V=W$ or $Z$) production in
association with one
or more jets ($V$+jets) is an important test of quantum chromodynamics
(QCD). In addition, $V$+jets processes are a significant background
to studies of Standard Model processes such as \ttbar~or
single-top production, as well as searches for the Higgs boson and for
physics beyond the
Standard Model.  Measurements of the cross section and kinematic
properties of $V$+jets processes and comparisons to theoretical
predictions are therefore of significant interest.
This Letter reports on a first measurement  at
the Large Hadron Collider (LHC) of the $W$+jets cross section 
in proton-proton ($pp$) collisions at a centre-of-mass energy
($\sqrt{s}$) of 7 \TeV, in both electron and muon decay modes of the
$W$-boson, with the ATLAS detector.  The measurement is
based on an integrated luminosity of approximately $1.3$ \ipb.

The cross section measurements are presented as a function of jet
multiplicity and of the transverse momentum (\pT) of the leading and
next-to-leading jets in each event.  Measurements are also presented
of the ratio of 
cross sections
$\sigma(W+ \ge n)/\sigma(W+ \ge n-1)$
for inclusive
jet multiplicities $n=1\rm{-}4$.
The results have been corrected for all known detector effects and are
quoted in a limited and well-defined  range of jet and lepton
kinematics, fully covered by the detector acceptance, so as to avoid
model-dependent extrapolations and to facilitate comparisons with
theoretical predictions.  Previous measurements of
$W$+jets production in proton-antiproton collisions at $\sqrt{s}=1.96$
\TeV~were published by the CDF Collaboration \cite{CDFWjets:2008}.
Theoretical calculations at next-to-leading-order (NLO) in
perturbative QCD (pQCD) have been computed
for up to four jets for $W$ production \cite{EllisWjets:2009,BlackHat4jets}.
Comparisons are made in this Letter
with NLO pQCD calculations for $n \le 2$;
higher jet multiplicities are compared only to leading-order (LO) calculations.

\section{The ATLAS Detector}

The ATLAS detector 
\cite{AtlasDetector,CSCbook} consists of an
inner tracking system (inner detector, or ID)
surrounded by a
thin superconducting solenoid providing a 2T magnetic field,
electromagnetic and hadronic
calorimeters and a muon spectrometer (MS).  The ID consists of pixel
and silicon microstrip (SCT) detectors, surrounded by the transition
radiation tracker (TRT).  The electromagnetic calorimeter is a
lead liquid-argon (LAr) detector.  Hadron calorimetry is based on two
different detector technologies, with scintillator-tiles or LAr 
as active media, and with either steel, copper, or tungsten as the
absorber material.
The MS is based on
three large superconducting toroids arranged with an eight-fold
azimuthal coil symmetry around the calorimeters, and a system of three
stations of chambers for the trigger and  for precise measurements.
The nominal $pp$ interaction point at the centre of the
detector
is defined as the origin
of a right-handed coordinate system.
The positive $x$-axis is defined by the direction
from the interaction point to the centre of the LHC ring, with the
positive $y$-axis pointing upwards, while the
beam direction defines the $z$-axis. The azimuthal angle $\phi$ is measured
around the beam axis and the polar angle $\theta$ is the angle from
the $z$-axis. The pseudorapidity is defined as $\eta = -\ln
\tan(\theta/2)$.

\section{Simulated Event Samples}

Simulated event samples were used for most of the background estimates,
for the correction of the signal yield for detector effects and for
comparisons of the results to theoretical expectations.
The detector simulation
\cite{AtlasSimulation}  was performed using GEANT4 \cite{Geant4}.
The simulated event samples are summarised in Table \ref{tab:MC}. 
The ALPGEN samples were generated with the MLM matching scheme
\cite{MLM} and
interfaced to HERWIG v6.510
\cite{Herwig} for parton shower and fragmentation processes and to
JIMMY v4.31 \cite{Jimmy} for underlying event simulation. 
Parton density functions (PDF) were: CTEQ6L1 \cite{CTEQ6L1} for the
ALPGEN and SHERPA samples, MRST 2007 $\rm{LO}^*$ \cite{LO*} for
PYTHIA, and MSTW2008 \cite{MSTW2008} for FEWZ \cite{FEWZ}.
For the POWHEG samples,
the PDF set was CTEQ6.6M \cite{CTEQ6.6M} for the NLO matrix element
calculations, while CTEQ6L1 was used for the parton showering and
underlying event via the POWHEG interface to PYTHIA.
The radiation of photons from charged leptons was treated in HERWIG and
PYTHIA using PHOTOS v2.15.4 \cite{Photos}. TAUOLA v1.0.2 \cite{Tauola}
was used for tau decays.
The underlying event tune was the ATLAS MC09 tune \cite{MC09tune} for
the ALPGEN samples, PYTHIA inclusive vector boson production, and
PYTHIA QCD samples.  The POWHEG sample used the ATLAS MC09 tune with
one parameter adjusted.\footnote{The cutoff for multiple parton
  interactions, PARP(82), was adjusted from 2.3
  to 2.1 GeV, suitable for the CTEQ6L1 PDF.}
The
AMBT1 \cite{AMBT1} tune was used for the PYTHIA $W$+jets samples.  The
samples generated with SHERPA used the default underlying event tune.
Samples were generated with
minimum bias interactions overlaid on top of the
hard-scattering event in order to account for the
multiple $pp$ interactions in
the same beam crossing (pile-up) experienced in the data.
The number of minimum bias interactions
followed a Poisson distribution with a mean of two \cite{PV}.
These samples were then reweighted such that
the distribution of the number of primary vertices matched that of the
data.

\begin{table*}[ht]
\begin{center}
\begin{tabular}{l|l|r r c}
\hline\hline
\raisebox{-0.4ex}{Physics process} & Generator&
\multicolumn{3}{c}{\raisebox{-0.4ex}{$\sigma \cdot \rm{BR}$  (nb)}} \\
\hline
\hline $W \rightarrow \ell \nu$ inclusive ($\ell = e,\mu,\tau$) &PYTHIA 6.4.21~\cite{Pythia}   & 10.46 &NNLO & \cite{FEWZ}  \\
 \hspace{0.5cm} $W^+\rightarrow \ell^+ \nu$        & & 6.16  &NNLO & \cite{FEWZ}  \\
 \hspace{0.5cm} $W^-\rightarrow \ell^- \overline{\nu}$ &         & 4.30  &NNLO & \cite{FEWZ}  \\
$W\rightarrow \ell \nu$ + jets ($\ell =
 e,\mu,\tau$) &PYTHIA 6.4.21~\cite{Pythia}   &   &  &   \\
$W\rightarrow \ell \nu$ + jets ($\ell = e,\mu,\tau$, $0 \le N_{parton} \le 5$) &ALPGEN 2.13~\cite{Alpgen}   &  &  &   \\
$W\rightarrow \ell \nu$ + jets ($\ell = e,\mu,\tau$,
$0 \le N_{parton} \le 4$) &SHERPA 1.1.3~\cite{Sherpa}   &   &  &   \\
 
\hline
$Z\rightarrow \ell \ell$ + jets ($m_{\ell\ell}>40$~GeV, $0 \le N_{parton} \le 5$) &ALPGEN 2.13~\cite{Alpgen} & 1.07 &NNLO&\cite{FEWZ} \\
$t\bar{t}$ &POWHEG-HVQ  & & & \\
           &  \hspace{0.5cm}v1.01 patch 4~\cite{Powheg} & 0.16 & NLO+NNLL  & \cite{Bonciani:1998vc}\\
Dijet ($e$ channel, $\hat{p}_{\mathrm{T}}>15$~GeV)      &PYTHIA 6.4.21~\cite{Pythia}  & 1.2$\times 10^6$  &LO&\cite{Pythia}\\
Dijet ($\mu$ channel,     $\hat{p}_{\mathrm{T}}> 8$~\GeV, $\pT^\mu>$~
8~\GeV)      &PYTHIA 6.4.21~\cite{Pythia}  & 10.6$\times 10^6$  &LO&\cite{Pythia} \\

\hline\hline

\end{tabular}
\caption{Signal and background simulated event
  samples used in this analysis, including the production cross
  section (multiplied by the relevant branching ratio, BR). The
  variable  $\hat{p}_{\mathrm{T}}$ is the average \pT~of the
  two outgoing partons involved in the hard-scattering process,
  evaluated before modifications from initial- and final-state
  radiation and from the underlying event. The $W$~inclusive
  cross section is given at next-to-next-to-leading-order
  (NNLO), the $t\bar{t}$ cross section is given at
  next-to-leading-order
  (plus next-to-next-to-leading-log, NNLL), and the dijet
  cross sections are given at leading-order (LO) in pQCD.  The
  $W$+jets and $Z$+jets samples were normalised using the inclusive
  cross sections.  For PYTHIA, the inclusive $W$ sample
  is based on a 
  $2\rightarrow 1$ matrix element merged with a  
  $2\rightarrow 2$ matrix element and a leading-logarithmic parton
  shower; the $W$+jets samples are based on  $2\rightarrow 2$ matrix
  elements. 
  Details of PDF sets,
  final-state photon radiation, and underlying event tunes are given in the text.}
\label{tab:MC}
\end{center}
\end{table*}

\section{Data and Event Selection}

The data used in this analysis
were collected from March to August 2010.  Application of
beam, detector, and data-quality requirements resulted in a total
integrated luminosity of 1.3 \ipb.
The uncertainty on the luminosity determination is
estimated to be 11\% \cite{lumi}.
Criteria for electron and muon
identification, as well as for event selection, followed closely those
for the $W$ boson inclusive cross section analysis \cite{incW}.

In the electron channel, a hardware-based level-one trigger system selected
events containing one or more electron candidates, based on the
presence of a cluster in the electromagnetic calorimeter with a
transverse energy (\ET) greater than 14 \GeV; this is the only
difference in the electron channel
with respect to the $W$~inclusive cross section analysis, and
was motivated by the fact that, for this larger dataset, this trigger
was the lowest-threshold, useful
electromagnetic trigger without any additional higher-level
trigger requirements.
The impact of the trigger efficiency was negligible for electrons with
\ET~$>$ 20 \GeV.
In the offline
analysis, electrons were required to pass the standard ``tight'' electron
selection criteria \cite{incW} with \ET~$>$ 20 \GeV~and
$|\eta| < 2.47$; electrons in the transition region
between the barrel and endcap calorimeter 
($1.37<|\eta|<1.52$) were rejected.  Events were also rejected if there was a
second electron passing the ``medium'' electron selection criteria
\cite{incW} and the same
kinematic selections as above.  

In the muon channel, the hardware-based
trigger selected events containing one or more muon
candidates, based on hit patterns in the MS,
corresponding to  \pT~$>$ 10 \GeV.
Offline, the muons were required to be identified in both ID and MS
subsystems and to have \pT~$>$ 20 \GeV~
and $|\eta| <
2.4$.
The ID track was
required to have $\ge 2$~hits in the pixel detector, $\ge 6$~hits
in the SCT and, for tracks with $|\eta| < 2.0$,
$\ge 1$~hit in the TRT.  The muon impact parameter with respect
to the primary vertex \cite{PV}
was required to be $<$~0.1~mm and $<$~10~mm in the
$r-\phi$ and $r-z$ planes, respectively.  The first of these
requirements
was added to further reduce non-prompt muons from decays of hadrons,
and muons from cosmic rays.
The difference between the ID
and MS \pT,
corrected for the mean energy loss in
upstream material, was required to satisfy
$|\pT^{\rm{ID}}-\pT^{\rm{MS}}| < 0.5\times \pT^{\rm{ID}}$.
Compared to the criteria used in Ref. \cite{incW}, this scaled
requirement reduced the background from decays-in-flight
of hadrons and improved the signal efficiency at high \pT.
As in Ref.~\cite{incW}, the muons
were required to be isolated, following a track-based isolation,
but the cone size was reduced from  $\Delta R = 0.4$ to $\Delta R =
0.2$ (where $\Delta R = \sqrt{(\Delta\eta)^2+(\Delta\phi)^2}$ of the muon)
and the isolation requirement was changed from $\Sigma
\pT^{\rm{ID}}/\pT < 0.2$ to $\Sigma \pT^{\rm{ID}} < 1.8$ \GeV~
to improve the QCD background rejection.
With these optimised cuts, the QCD background was reduced by
a factor of 1.7 for the inclusive 1-jet sample.
In addition, a number of requirements were added on the tracks inside the
isolation cone: the difference between the $z$ position of the track
extrapolated to the beam line and the $z$ coordinate of the primary
vertex was required to be $< 1$~cm, and the total number of hits
in the pixel and SCT detectors was required to be $\ge 4$.  These
additional requirements further improved the rejection of QCD
background. 
Events were rejected if there was a second muon passing
the same kinematic selections and isolation requirements as above.

The calculation of missing
transverse energy (\met) and transverse mass ($M_{\rm{T}}$)
followed the prescription in Ref.~\cite{incW}.  $M_{\rm T}$ was
defined by the lepton and neutrino \pT~as
$\ensuremath{M_{\mathrm{T}}} =
\sqrt{2\ensuremath{p_{\mathrm{T}}}^{\ell}\ensuremath{p_{\mathrm{T}}}^{\nu}(1-\cos(\phi^{\ell}-\phi^{\nu}))}$,
where the $(x,y)$ components of the neutrino momentum were inferred
from the corresponding \met~components.
\met~was calculated from
the energy deposits of calorimeter cells inside
three-dimensional clusters \cite{topoclusters}.  These clusters were
then corrected to take into account the different response to hadrons
compared to electrons or photons, as well as dead material and
out-of-cluster energy losses \cite{lochadtopo}.  In the muon channel,
\met~was corrected for the muon momentum.  Events were required to have
\met~$>$ 25 GeV and $M_{\rm{T}} > 40$~\GeV.
After requiring $\ge 1$~primary vertex with $\ge 3$ associated tracks
in the event,
the primary vertex was
required to be within 150~mm along the beam
direction relative to the centre of the detector. In events with
multiple vertices along the beam axis, the vertex with the largest
$\Sigma \pT^2$
 of associated
tracks was taken as the primary event vertex.
Starting from approximately $9.6 \times10^{6}$  triggered
events in each of the electron and muon channels, these
selection criteria reduced the sample to 4216 and 4911 events, respectively.

Jets were reconstructed using the anti-$k_{\rm{t}}$ algorithm
\cite{Cacciari:2008gp} with a radius parameter $R = $ 0.4 \cite{JES}.  The
efficiency for reconstructing jets was found to be approximately 98\%
in simulation for jet \pT~of 20 GeV, rising to close to 100\%
efficiency for 30 GeV jets. Jets
arising from detector noise or cosmic rays were
rejected \cite{jetcleaning}.
To take into account the differences in
calorimeter response
to electrons and hadrons, a \pT- and $\eta$-dependent factor,
derived from simulated events, was applied to each jet to provide an average
energy scale correction \cite{JES} back to particle-level.
Jets were required to
have $|\eta| < 2.8$ and \pT~$>$ 20 \GeV.  All jets within
$\Delta R < 0.5$ 
of an electron or muon (that passed the lepton identification
requirements) were removed,
regardless of the jet \pT~or $\eta$.

Jets from pile-up interactions were removed by a cut on
the jet-vertex
fraction ($JVF$) which was computed for each jet in the event.
After associating tracks to jets with a simple
matching in $\Delta R$(track,~jet), requiring $\Delta R < 0.4$,   the
$JVF$ was computed for each jet
as the scalar sum \pT~of all matched tracks from the primary
vertex divided by the total jet-matched track \pT~from all vertices.
 Jets which fell outside of the fiducial tracking
region ($|\eta| < 2.5$) or which had no matching tracks were not considered for the $JVF$
cut.  Jets for which $JVF$ $<$ 0.75   were rejected.  The application of
the $JVF$ cut reduced the sensitivity of the measured jet multiplicity
distribution to additional jets from
pile-up.

\section{Signal and Background Yields}

The major background processes in the electron channel are 
QCD and leptonic
backgrounds.  The latter consist of  
\Wtau~where the tau decays to an electron,
\Zee~where one electron is not identified and hadronic energy in the
event is mismeasured,  and
semileptonic \ttbar~decays ($\ttbar \rightarrow b\overline{b} q q' e
\nu$). 
The QCD
background in the electron channel has two components, one where a hadronic jet
passes the electron selection and
additional energy mismeasurement in the event results in large \met,
and the other where a bottom- or charm-hadron decays to an electron. 
For the muon channel, the main backgrounds arise from semileptonic
heavy flavour decays in multijet events and from the leptonic
background from the following sources: \Wtau~where the tau
decays to a muon, \Zmm~where one muon is not identified, and
semileptonic \ttbar~decays in the muon channel.  
The contributions of single-top and diboson production to the measured
cross section have been estimated to be slightly smaller than the
\Wtau~background, and are not subtracted from the data.

\begin{figure*}[t]
  \begin{minipage}{0.5\linewidth}
  \begin{center}
    \includegraphics[width=0.95\linewidth]{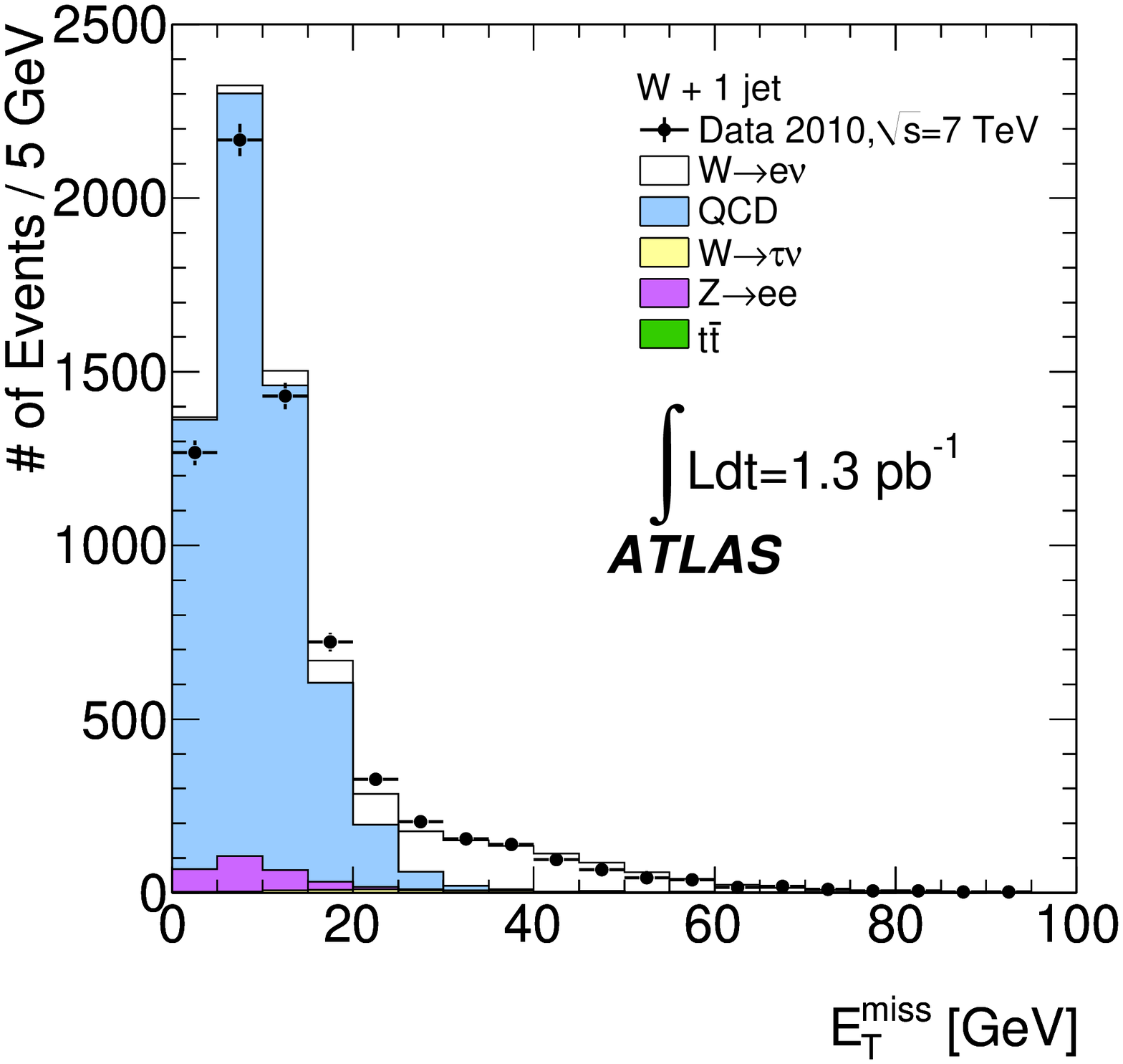}
  \end{center}
  \end{minipage}
  \begin{minipage}{0.5\linewidth}
  \begin{center}
    \includegraphics[width=0.95\linewidth]{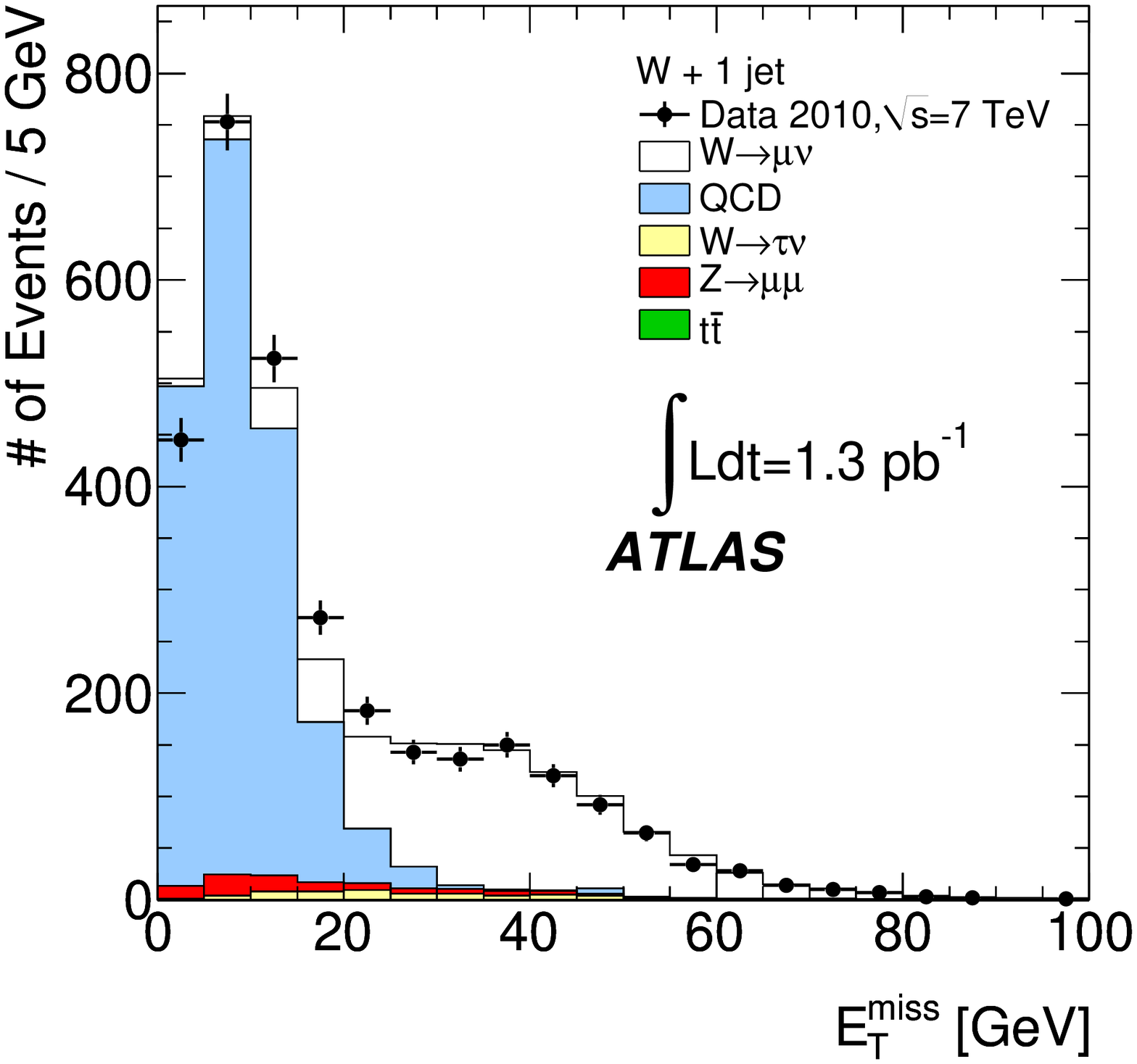}
  \end{center}
  \end{minipage}
  \caption{Results of fitting the signal and background templates to
    the \met~distribution in the electron (left) and muon (right)
    channels for the 1-jet bin.  In the electron channel, the fit was
    performed for \met~$>$ 10 GeV. All templates were from simulated
    events, except for the QCD background template in the  electron
    channel which was obtained from the data.}
  \label{fig:template_fit}
\end{figure*}

The number of leptonic background events surviving the above selection
cuts was estimated with simulated event samples: ALPGEN for vector boson
samples (PYTHIA was used for $W \rightarrow \tau \nu$~+ jets)
and POWHEG for \ttbar~background.  
The simulated leptonic background samples were
normalised to the integrated luminosity of the data using the
predicted NNLO or NLO+NNLL cross sections.
The number of QCD background events was estimated by fitting, in
each jet multiplicity bin, the
\met~distribution in the data (without the \met~and $M_{\rm{T}}$ cuts)
to a sum of two templates: one for the QCD background and
another which included signal and the leptonic backgrounds. 
In both muon and electron channels, the shapes for the second
template were obtained
from simulation.  In the electron channel, the template for the QCD
background was obtained from the data because the mechanisms by
which a jet fakes an electron
are difficult to simulate.  This template was derived from
a data sample where looser electron identification criteria were applied on the
shower shapes and the track-cluster matching requirements were inverted.
The QCD background was computed from the results of the template fit.
In the electron channel, the fit was performed in the region $\met~>
10$ \GeV~due to the poor understanding of the background below 10~\GeV.
The fit to the \met~distribution was used only to determine the QCD background
normalisation, taking into account contributions from leptonic
background and signal in the low \met~region.
The $W$+jet signal yield for the cross section calculation was derived as
the difference between the observed number of events in the signal
region and the sum of background components.
Figure \ref{fig:template_fit} shows
the \met~distribution for events with one jet, with the
fitted contributions from all background sources in the electron and
muon channels respectively, 
after all the
other selection requirements (except for the $M_{\rm{T}}$ cut)
have been applied.  The residual difference in \met~between the data and the
QCD template in the control region is covered by the systematic uncertainties.
The number of observed events and the 
estimated number of background events are summarised in
Table \ref{tab:sigbkg_summary}.

\begin{table*}[ht]
\begin{center} 
\begin{tabular}{l|c|c|c|c|c} \hline\hline
Electron channel  &              &              &             &          &\\
process           & $N_{\rm{jet}} \ge 0$ & $N_{\rm{jet}} \ge 1$ &
$N_{\rm{jet}} \ge 2$ & $N_{\rm{jet}} \ge 3$ & $N_{\rm{jet}} \ge 4$ \\ \hline
&&&&&\\
QCD   & $130 \:^{+20}_{-60}$ &
      $ 100 \:^{+20}_{-40}$ &
      $ 45 \:^{+7}_{-20} $ &
      $ 18 \:^{+3}_{-8}  $ & - \\
\Wtau & $113 \pm 11  $ &
        $25 \pm 5 $ &
        $ 4 \pm 2 $ &
        $ 0.5 \pm 0.4  $ & -  \\
\Zee &  $ 10 \pm 8 $ &
        $ 7 \pm 6 $ &
        $ 3 \pm 2 $ &
        $ 1 \pm 1  $ & -  \\
\ttbar &$17 \pm 2 $ &
        $17 \pm 2 $ &
        $17 \pm 2 $ &
        $14 \pm 2 $ & -  \\ \hline 

Observed in Data & 4216 & 987 & 276 & 83 & - \\ \hline\hline

\multicolumn{6}{c}{  } \\ \hline\hline
Muon channel      &              &              &             &          &\\
process           & $N_{\rm{jet}} \ge 0$ & $N_{\rm{jet}} \ge 1$ &
$N_{\rm{jet}} \ge 2$ & $N_{\rm{jet}} \ge 3$
&$N_{\rm{jet}} \ge 4$ \\ \hline
&&&&&\\
QCD   &  $30 \pm 20$  &
                $20 \pm 13$  &
                $4\:^{+10}_{-4}$  &
                $2\pm 2$  &
                $1\pm 1$  \\

\Wtau &  $133 \pm 12 $  &
         $24 \pm 6 $  &
         $5 \pm 2 $  &
         $0.9 \pm 0.5$  &
         $0.4\pm 0.3$  \\

\Zmm &  $170\pm 14$  &
        $30\pm 4$  &
        $8 \pm 1 $  &
        $2\pm 0.5$  &
        $0.6\pm 0.2 $  \\

\ttbar &  $18 \pm 2$  &
          $18 \pm 2$  &
          $18 \pm 2$  &
          $16 \pm 2$  &
          $11 \pm 1$  \\\hline

Observed in Data & 4911 & 1049 & 292 & 95 & 36 \\ \hline\hline

\end{tabular}
\caption{Summary of background
  yields and observed number of events
  for the electron and muon channels with systematic
  uncertainties, excluding the luminosity uncertainty.
  Statistical uncertainties are
  negligible compared to systematic uncertainties.
  The uncertainty in the backgrounds due to the
  luminosity uncertainty is 11\% for all backgrounds
  except for the QCD background, since it was normalised to the
  data.
  The measurement was not performed in the inclusive 4-jet
bin in the electron channel because of the poor signal-to-background
ratio. }
\label{tab:sigbkg_summary}
\end{center} 
\end{table*}

The yield of signal events was corrected back to the particle level,
taking into account detector and
reconstruction efficiency.
The dominant corrections
in the electron channel come from electron
reconstruction efficiency ($\approx 20$\% correction).
In the muon channel, the dominant corrections come from trigger
and reconstruction efficiency (corrections of $\approx 10-20$\% and $\approx
10$\% respectively).
The corrections were computed using
the ALPGEN $W$+jets event generator plus full detector simulation,
restricting the events to the same phase space as the data analysis.  The phase
space requirements were applied to generated quantities.  In this analysis,
particle-level jets
were constructed in simulated events
by applying the jet finder
to all final state  particles (excluding muons and neutrinos)
with a lifetime longer than 10~ps,
whether produced directly in the $pp$ collision or from the decay
of particles with shorter lifetimes.
Correction factors were computed as one-dimensional
functions of jet multiplicity and \pT~of the leading and next-to-leading
jets, and were treated as independent.
Migration of events across bins of jet \pT~was
made small compared to the statistical uncertainty by selecting
the bin widths to be at least a factor of two larger than the jet
\pT~resolution \cite{JES}. Tests with simulated data showed that
these correction factors were sufficient to recover
particle-level distributions.
To  treat the effect
of final state QED radiation, the energy of the generated lepton was
defined as the energy of the lepton after radiation plus the energy of
all radiated photons within $\Delta R=0.1$ around the
lepton.

The correction factor for the trigger efficiency was obtained
directly from the data as follows.
In the electron channel, events were triggered
either by an independent \met~trigger or a loose electron trigger
with an approximately 5
\GeV~threshold.  The full $W$+jets selection was carried out in
essentially the same way as described above
in order to isolate a pure electron sample.
The main difference was in the QCD background estimation,
which was done with templates for the shape of the electron isolation
distribution, where the isolation variable was defined as the sum of
transverse energy in a  cone of $\Delta R=0.4$ around the electron divided by
the transverse energy of the electron.
These templates were obtained by inverting one or more
of the electron shower shape requirements.  The electron trigger efficiency
was found to be close to 100\% in both data and simulation.  In the
muon channel, the trigger efficiency was computed with a sample of unbiased
offline reconstructed muons from \Zmm~decays.  Average trigger
efficiencies of $82.0 \pm 1.4$\% and $86.9\pm0.1$\% were obtained in data
and simulation, respectively; the difference between data and
simulation comes from a mismodelling of both the efficiency of the forward
muon chambers and of the programming of the muon trigger
electronics.  The trigger efficiency (and its uncertainty)
from the data was used for the correction factor.

\section{Systematic Uncertainties}

The primary sources of systematic uncertainty in the cross section for
both electron and muon channels are uncertainties in the integrated
luminosity and in the jet energy
scale \cite{JES}.    In the electron channel, the uncertainty
due to the QCD background shape is also important. Both electron and
muon channels
are affected by uncertainties in
the lepton reconstruction efficiency.
The luminosity uncertainty enters primarily through the signal
normalisation but also has an effect on
the estimation of the leptonic backgrounds; the luminosity uncertainty
is therefore larger in the muon channel.

Uncertainties in the jet energy scale (JES)
and jet energy resolution (JER) were determined primarily from
simulations \cite{JES}. The JES
uncertainty varies as a function of jet \pT~and $\eta$, and ranges
from around 10\% at 20 \GeV~to about 8\% at 100 GeV.
The JER uncertainty is 14\% of the jet energy
resolution.
To take into account the differences in calorimeter response to quark-
and gluon-initiated jets, an additional uncertainty
of 5\% was added in quadrature to the JES
uncertainty, based on the
average difference in simulation of the calorimeter response between jets in the
$W$+jets samples compared to those in the dijet samples (on which the JES
calibration is based). Uncertainties in the JES due to nearby jets
in $W$+jets events were also studied but found to be small.
To estimate the impact of the JES uncertainty,
jet energies in the simulated events were shifted  by the JES
uncertainty, and the \met~vector was recomputed.
In addition, calorimeter clusters not
associated to a jet or electron, such as those coming from the underlying
event, were scaled using a \pT-dependent
uncertainty \cite{incW}, ranging from $\pm
20$\% for \pT~$\simeq 500$~\MeV~to $\pm 5$\% at high \pT.
Similarly the jet energies were smeared with a Gaussian representing
the JER uncertainty and the \met~vector was recomputed.  
The full analysis was repeated with these variations,
and the cross sections were
recomputed;  the change in the cross section
was taken as the systematic uncertainty. The impact of the JES and
\met~uncertainties on the cross section uncertainty was
approximately 10\%.

A significant source of uncertainty in the electron channel is
the potential bias in the sample selection for
building the template shape of the QCD background; with the current selection
requirements, the contribution from semileptonic heavy flavour decays
is underestimated. The size of the effect was
determined with simulated events by comparing the background estimates
from two templates: one based on
the electron selection used  for this cross section measurement and the
other based on the selection used for the QCD background estimation in
the electron channel.  The
resulting uncertainty on the QCD background estimate, including
significant contributions from the limited statistics of the simulated
event samples, was as high as 50\%, but the effect on the cross
section for the inclusive 1-jet bin was about 5\%.
The fit region for the QCD background
was varied by $\pm 5$~\GeV~to account for shape
differences in the low \met~region; the resulting uncertainty on the
cross section was $1-2$\%.

The uncertainty in the electron identification efficiency was taken from
the inclusive cross section measurement \cite{incW}.  By
examining the reconstruction efficiency in simulated
events as a function of the $\Delta R$ separation between the jet and
the electron,  the reconstruction efficiency was
found to be consistent with the value in Ref.~\cite{incW}.
Furthermore, in the region $\Delta R > 0.5$, the efficiency was
found to be  constant as a function of $\Delta R$ and as a
function of jet multiplicity.
The uncertainty in the muon
reconstruction efficiency was estimated by comparing the efficiency
measured with simulated events to that measured in the data with
muons from \Zmm~decays, following a method similar to that described
in Ref.~\cite{incW}.  The resulting uncertainties in the cross
section were approximately 5.5\% in both electron and muon channels.

Other uncertainties which were considered include the trigger
efficiency, jet reconstruction efficiency,
lepton momentum scale and resolution, pile-up, and biases
in the procedure for correcting for detector effects (for example, by
comparing correction factors  obtained with ALPGEN to those obtained
with SHERPA).
Their effect on the cross section was found to be smaller
than the uncertainties described above.  For example, the
uncertainty on the electron energy resolution was based on
extrapolations from test-beam measurements \cite{incW}
and had a $< 0.1$\% effect
on the cross section.
All of the above systematic uncertainties
(except for the bias in the template shape for the QCD background in
the electron channel) were also applied to
the estimates of the QCD and leptonic backgrounds in both electron and
muon channels. In addition, for the leptonic backgrounds
the uncertainty in the
NNLO cross sections was taken to be 5\% for $W/Z$ production as in
Ref.~\cite{incW}. The \ttbar~cross section uncertainty was taken to be
approximately 7\%, amounting to the sum in quadrature of PDF uncertainties (3\%)
and uncertainties estimated by varying renormalisation and
factorisation scales (6\%) \cite{Moch:2008qy,Beneke:2009ye}.

\begin{figure*}[htbp]
  \begin{minipage}{0.5\linewidth}
  \begin{center}
    \includegraphics[width=0.95\linewidth]{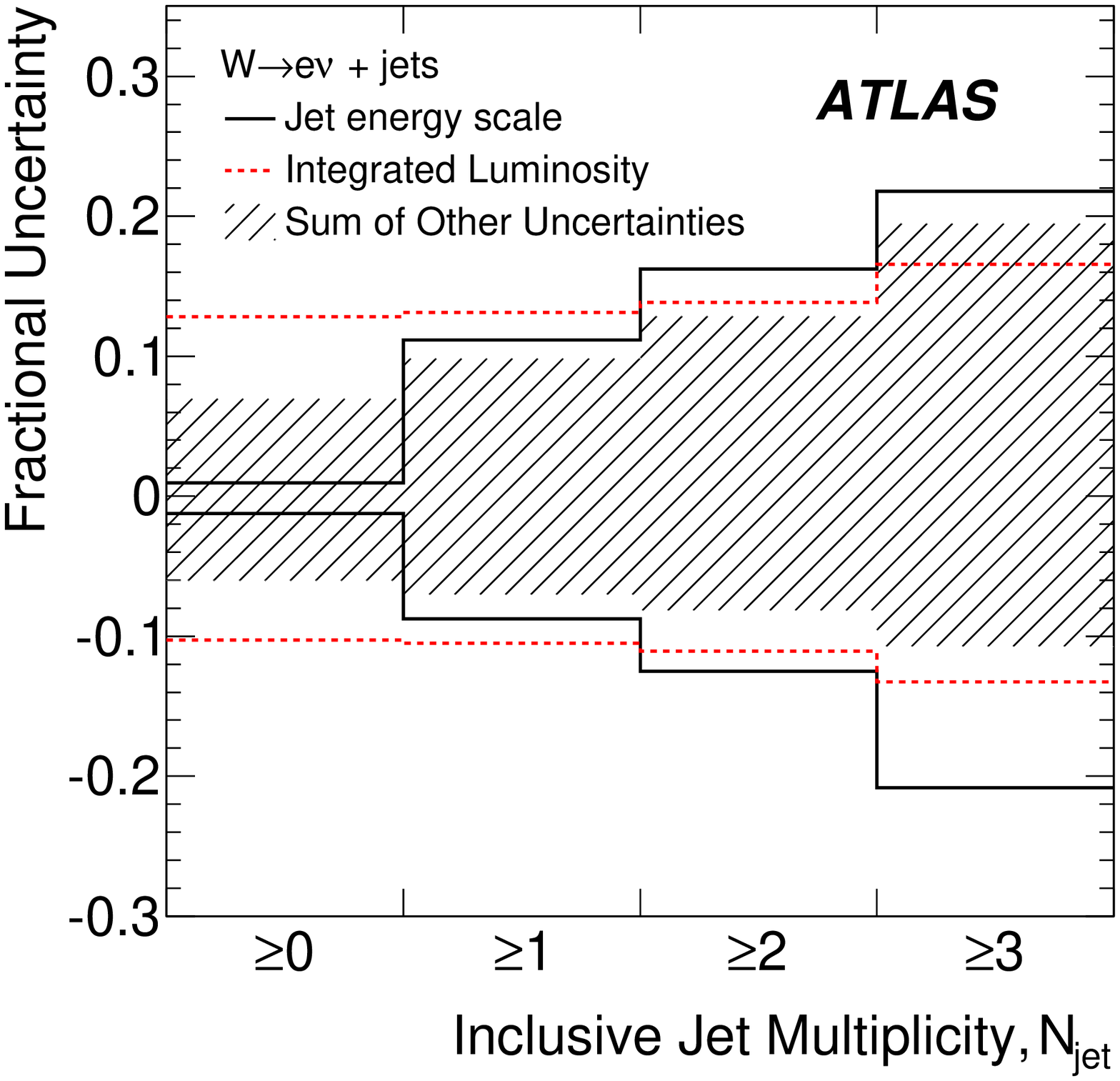}
  \end{center}
  \end{minipage}
  \begin{minipage}{0.5\linewidth}
  \begin{center}
   \includegraphics[width=0.95\linewidth]{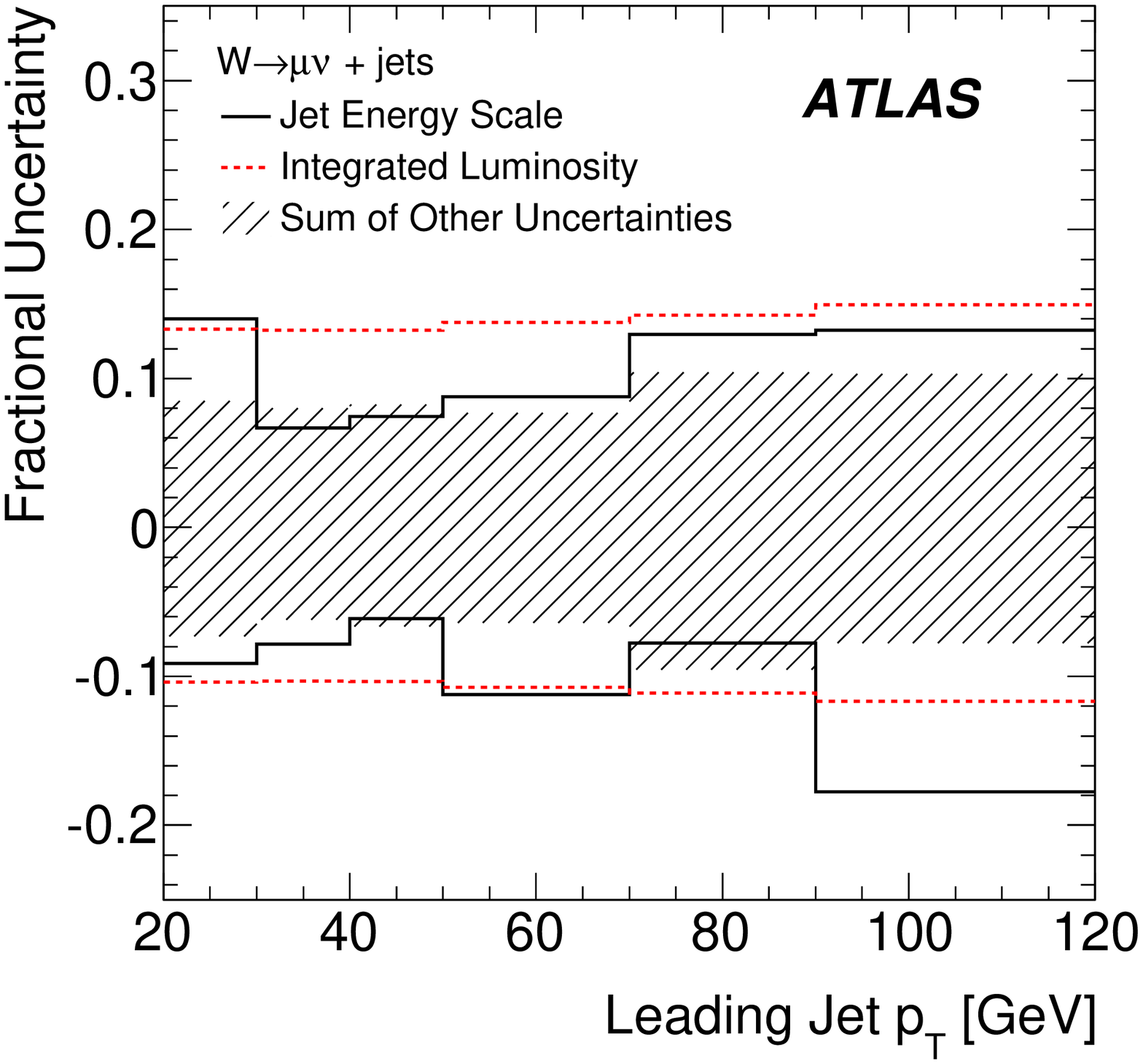}
  \end{center}
  \end{minipage}
  \caption{Summary of the systematic uncertainties on the cross
    section measurement shown as a 
    function of jet multiplicity in the electron channel (left) and
    leading-jet \pT~in the muon channel (right). The jet energy scale
    uncertainty includes the uncertainty on \met.
The main contribution to the
``sum of other uncertainties'' in the electron channel
comes from the QCD background (especially at high jet multiplicities),
the electron identification efficiency and the electron energy scale.
For the muon channel, the main contribution is from the muon
reconstruction efficiency.}
  \label{fig:syserr}
\end{figure*}

The systematic uncertainties in the cross section measurement are
summarised in Table \ref{tab:effsys_summary} for $N_{\rm{jet}} \ge 1$;
most of the uncertainties are approximately independent of jet
multiplicity, except for the uncertainty due to the jet energy scale
and resolution, and the QCD background in the electron channel.
The dominant systematic uncertainties are shown as a function
of jet multiplicity and leading jet \pT~in Figure \ref{fig:syserr}.
Both distributions are similar for electron and muon channels; the
uncertainty is therefore shown as a function of jet multiplicity for the
electron channel and as a function of leading jet \pT~for the muon
channel.  The main contribution to the
other uncertainties in the electron channel
comes from the QCD background (especially at high jet multiplicities),
the electron identification efficiency and the electron energy scale.
For the muon channel, the main contribution is from the muon
reconstruction efficiency.

\begin{table*}[ht] 
\begin{center} 
\begin{tabular}{l|c|c} \hline\hline
\multicolumn{3}{c}{$e$ channel} \\\hline
         &       & Cross Section \\
  Effect & Range & Uncertainty (\%) \\\hline
  Jet energy scale and \met& $\pm10$\% (dependent on jet $\eta$ and \pT)~
  $\oplus$ 5\% & +11, -9 \\
  Jet energy resolution & 14\% on each jet & $\pm1.0$ \\
  Electron trigger & $\pm0.5$\% & $\mp0.7$ \\
  Electron identification & $\pm5.2$\% & $\mp5.5$ \\
  Electron energy scale & $\pm$3\%  &
  +3.9, -4.7 \\
  Pile--up removal cut & $4-7$\% in lowest jet \pT~bin & $\pm1.9$ \\
  Residual pile-up effects & from simulation  & $\pm2.2$ \\
  QCD background shape & from template variation &
  -1.5, +5.2 \\
  Luminosity & $\pm11$\% & -10, +13 \\\hline\hline
  \multicolumn{3}{c}{    } \\\hline\hline
  \multicolumn{3}{c}{$\mu$ channel} \\\hline
         &       & Cross Section \\
  Effect & Range & Uncertainty (\%) \\\hline
  Jet energy scale and \met & $\pm10$\% (dependent on jet $\eta$ and \pT)~
  $\oplus$ 5\% & +11, -9 \\
  Jet energy resolution & 14\% on each jet &  $\pm1.8$ \\
  Muon trigger & $\pm2.5$\% in barrel, $\pm2.0$\% in endcap & $\mp1.6$ \\
  Muon reconstruction & $\pm5.6$\% &  -5.4, +5.9  \\
  Muon momentum scale & $\pm1$\%  & +2, -0.9   \\
  Muon momentum resolution & $\pm5$\% in barrel, $\pm9$\% in endcap  & $\pm1.4$  \\
  Pile--up removal cut & $4-7$\% in lowest jet \pT~bin & $\pm1.7$ \\
  Residual pile-up effects & from simulation  & $\pm1.4$ \\
  Luminosity & $\pm11$\% & -11, +13 \\\hline\hline
\end{tabular}
\caption{Summary of the systematic uncertainties in the cross section.
  The uncertainties are
  shown only for $N_{\rm{jet}} \ge 1$.  The sign
  convention for the JES and lepton energy scale uncertainties is such
that a positive change in the energy scale results in an increase in
the jet or lepton energy observed in the data. }
\label{tab:effsys_summary}
\end{center} 
\end{table*}

In the cross section ratio measurement, the uncertainty due to the jet
energy scale uncertainty remains the dominant effect, amounting to
approximately 10\% on the ratio.  The luminosity
uncertainty does not completely cancel in the ratio because the background
estimates are affected by the luminosity uncertainty and the
background levels vary as a function of jet multiplicity.

\section{Results and Conclusions}

\begin{figure*}
  \begin{minipage}{0.5\linewidth}
  \begin{center}
    \includegraphics[width=0.95\linewidth]{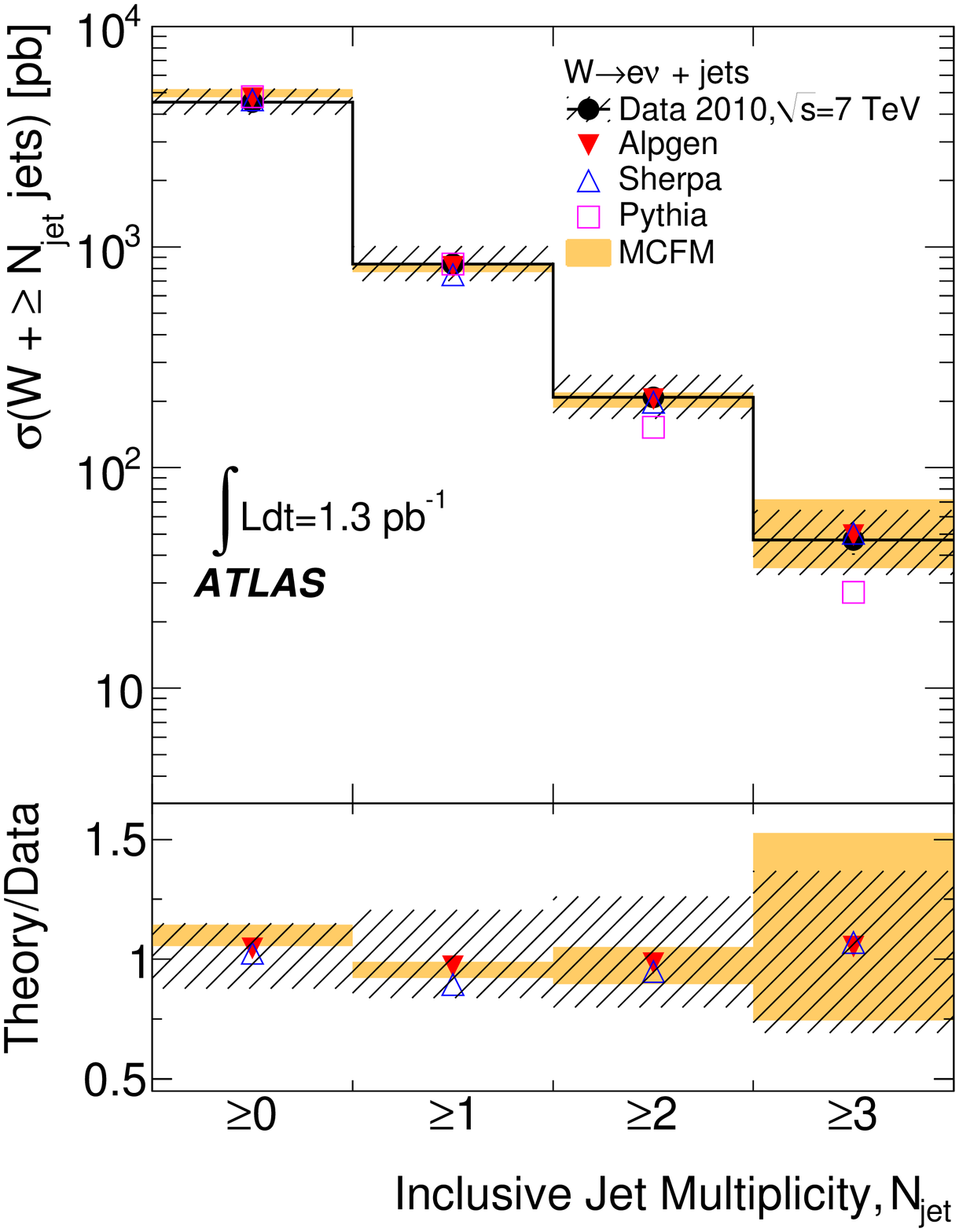}
  \end{center}
  \end{minipage}
  \begin{minipage}{0.5\linewidth}
  \begin{center}
    \includegraphics[width=0.95\linewidth]{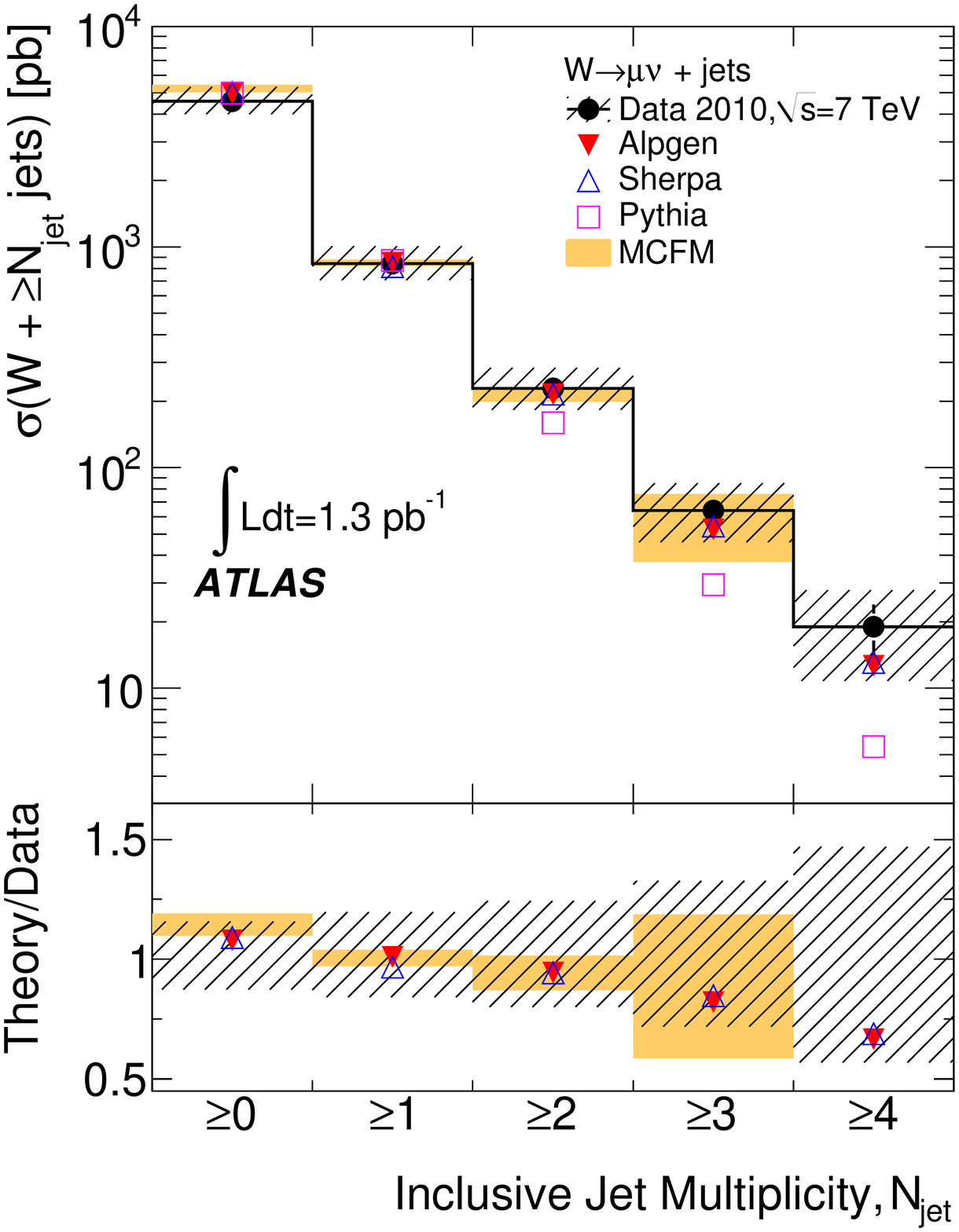}
  \end{center}
  \end{minipage}
  \caption{$W$+jets cross-section results as a function of corrected
    jet multiplicity.
    Left: electron channel.  Right: muon channel.  The cross
    sections are quoted in a limited and well-defined
    kinematic region, described
    in the text.  For the data, the statistical uncertainties
    are shown by the vertical bars, and the combined statistical
    and systematic uncertainties are shown by the hashed regions.
    Note that the uncertainties are correlated from bin to bin.
    Also shown are  predictions
    from PYTHIA, ALPGEN, SHERPA
    and MCFM, and the ratio of theoretical predictions to data (PYTHIA
    is not shown in the ratio). The theoretical
    uncertainties are shown only for MCFM, which provides NLO predictions
    for $N_{\rm{jet}} \le 2$ and a LO prediction for $N_{\rm{jet}} = 3$.}
  \label{fig:result-xsec}
\end{figure*}

\begin{figure*}
  \begin{minipage}{0.5\linewidth}
  \begin{center}
    \includegraphics[width=0.95\linewidth]{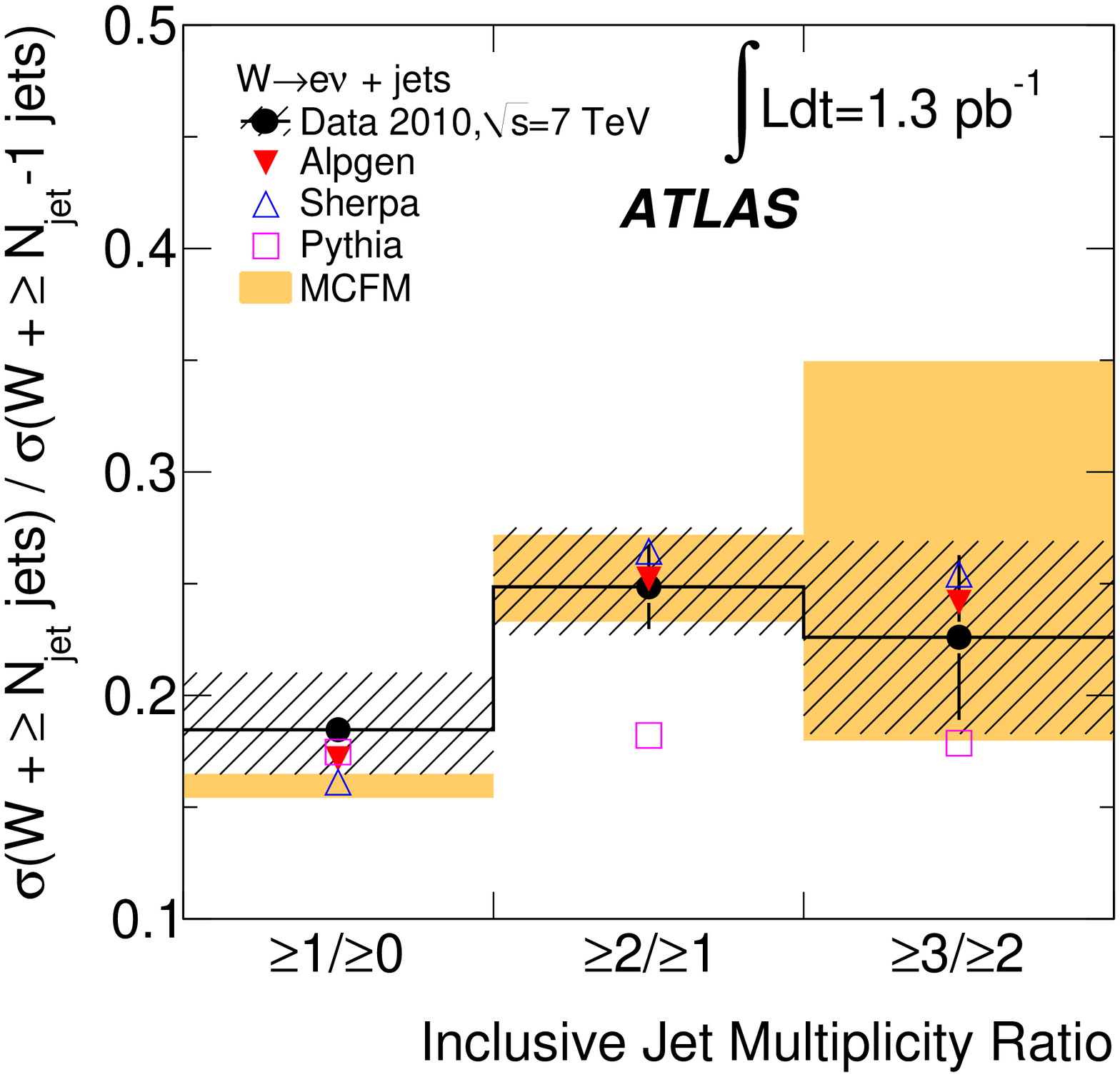}
  \end{center}
  \end{minipage}
  \begin{minipage}{0.5\linewidth}
  \begin{center}
    \includegraphics[width=0.95\linewidth]{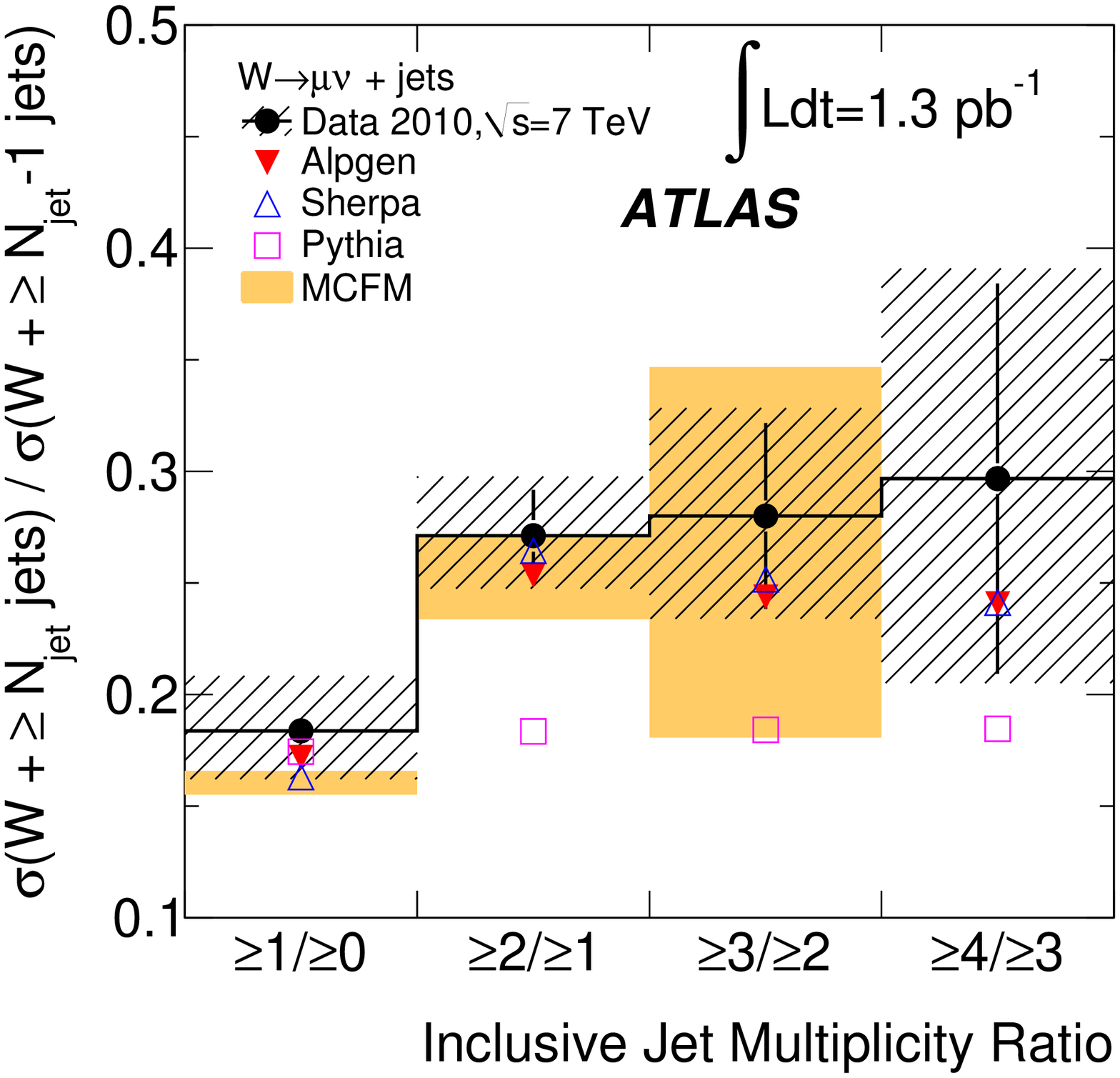}
  \end{center}
  \end{minipage}
  \caption{$W$+jets cross-section ratio results as a function of
    corrected jet multiplicity.
    Left: electron channel.  Right: muon channel.  The cross
    sections are quoted in a limited and well-defined
    kinematic region, described
    in the text.   For the data, the statistical uncertainties
    are shown by the vertical bars, and the combined statistical
    and systematic uncertainties are shown by the hashed regions.
    Also shown are theoretical predictions
    from PYTHIA, ALPGEN, SHERPA,
    and MCFM. The theoretical
    uncertainties are shown only for MCFM, which provides NLO predictions
    for $N_{\rm{jet}} \le 2$ and a LO prediction for $N_{\rm{jet}} = 3$.}
  \label{fig:result-xsecratio}
\end{figure*}

\begin{figure*}
  \begin{minipage}{0.5\linewidth}
  \begin{center}
    \includegraphics[width=0.95\linewidth]{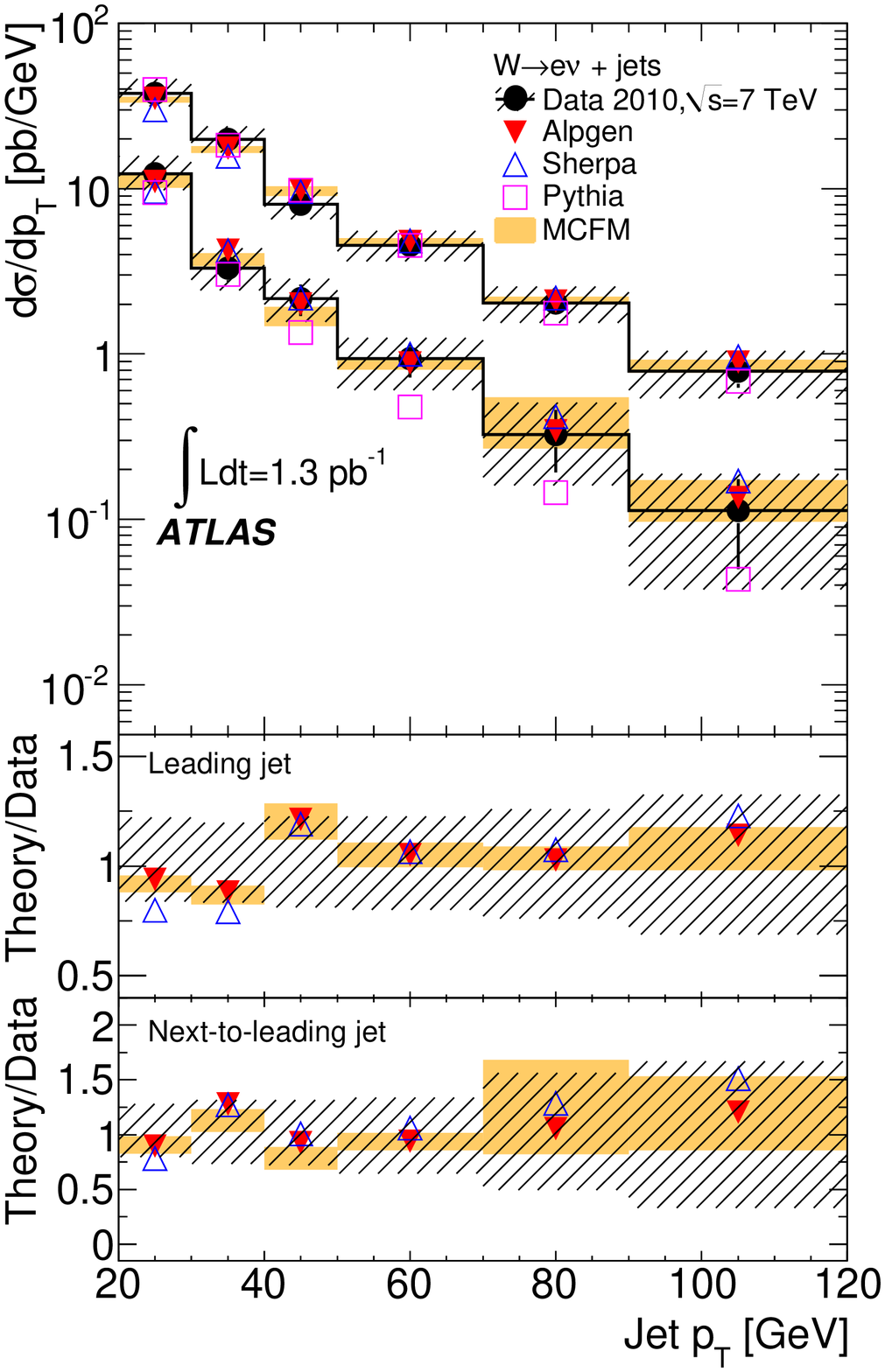}
  \end{center}
  \end{minipage}
  \begin{minipage}{0.5\linewidth}
  \begin{center}
    \includegraphics[width=0.95\linewidth]{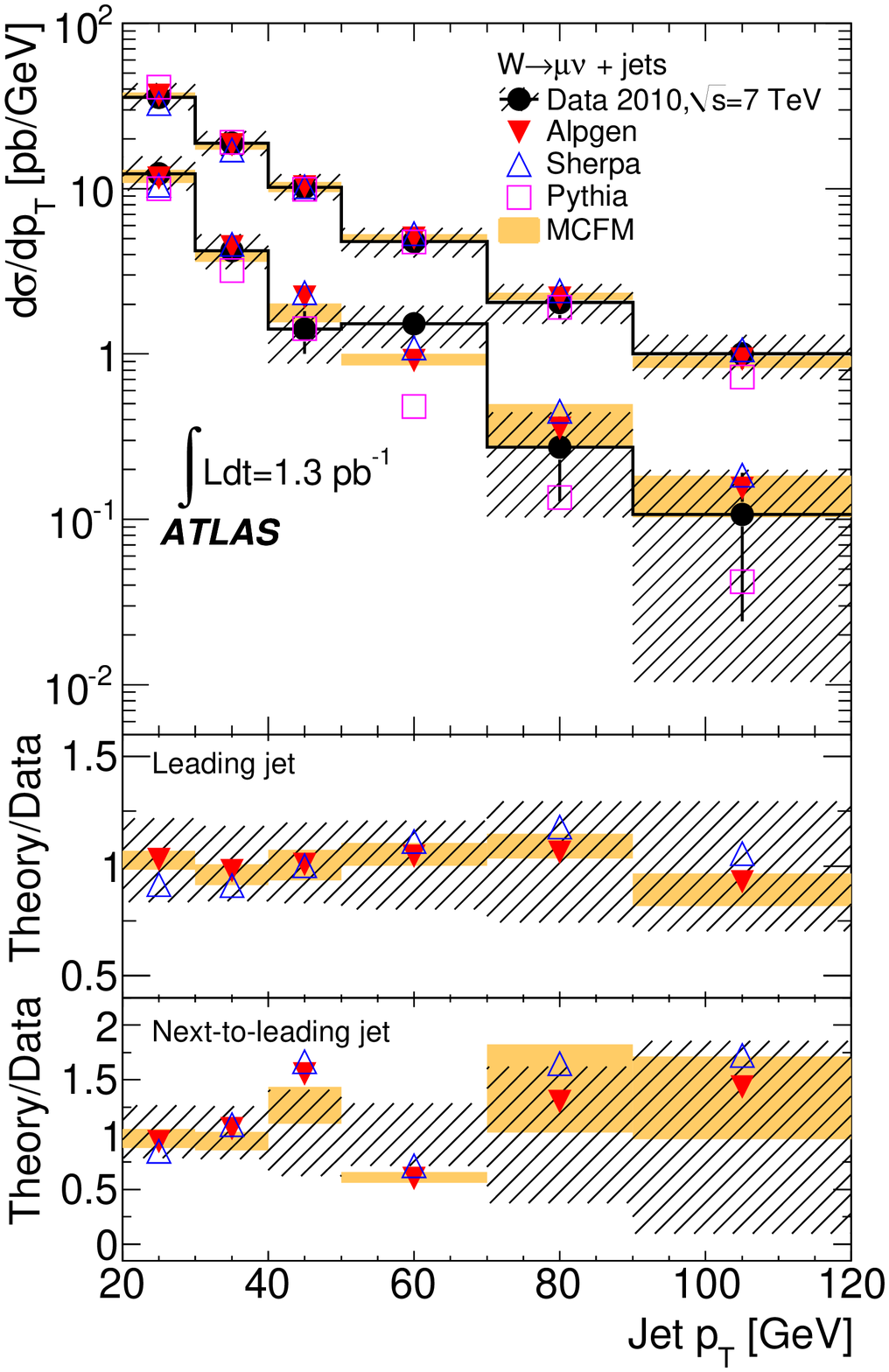}
  \end{center}
  \end{minipage}
  \caption{$W$+jets cross-section as a function of the \pT~of the
    two leading jets in the event.  The \pT~of the leading jet is
    shown for events with $\ge 1$~jet while the \pT~of the
    next-to-leading jet is shown for events with $\ge 2$~jets.
    Left: electron channel.  Right: muon channel.  The cross
    sections are quoted in a limited and well-defined
    kinematic region, described
    in the text.   For the data, the statistical uncertainties
    are shown by the vertical bars, and the combined statistical
    and systematic uncertainties are shown by the hashed regions.
    Also shown are theoretical predictions
    from PYTHIA, ALPGEN, SHERPA
    and MCFM, and the ratio of theoretical predictions to data (PYTHIA
    is not shown in the ratio). The theoretical
    uncertainties are shown only for MCFM, which provides NLO predictions
    for $N_{\rm{jet}} \le 2$ and a LO prediction for $N_{\rm{jet}} = 3$.}
  \label{fig:result-xsec-jetpt}
\end{figure*}

The measured $W$+jets cross section (multiplied by the leptonic
branching ratio) and the cross section ratios are shown as a function
of corrected jet multiplicity
in Tables \ref{tab:xsec_vs_multiplicity} and
\ref{tab:xsec_ratio} respectively,
as well as in Figures \ref{fig:result-xsec} and
\ref{fig:result-xsecratio}.
The measurement was not performed in the inclusive 4-jet bin in the
electron channel because of the poor signal-to-background ratio.
The cross
sections are quoted in the limited kinematic region: $\ET^{\rm{j}} >$
20 \GeV, $|\eta^{\rm{j}}| < 2.8$, $\ET^\ell >$ 20 \GeV,
$|\eta^{\rm{e}}| < 2.47$ (excluding $1.37<|\eta^{\rm{e}}|<1.52$),
$|\eta^{\mu}| <
2.4$, $\pT^{\nu} >$ 25 \GeV, $M_{\rm{T}} > 40$~\GeV, $\Delta R^{\rm{lj}}
> 0.5$, where $\ell$, j and $\nu$ denote lepton, jet and neutrino,
respectively.   
The quantities $\pT^\ell$, $|\eta^\ell|$, and
$M_{\rm{T}}$ include the energy of
all radiated photons within $\Delta R=0.1$ around the
lepton.  
The $W$+jets
cross section (times leptonic branching ratio) is shown as a function
of the \pT~of the leading and next-to-leading jets in the event
in Figure \ref{fig:result-xsec-jetpt}; the leading jet is shown for
$N_{\rm{jet}} \ge 1$ and the next-to-leading jet is shown for
$N_{\rm{jet}} \ge 2$.

\begin{table*}[ht]
  \begin{center}
    \begin{tabular}{c|c|c|c|c}
      \hline \hline
        Jet                 &           & MCFM           &     & MCFM \\
      multiplicity      & \Wen~(nb) & \Wen~(nb) & \Wmn~(nb) &
      \Wmn~(nb) \\ 
      \hline
      &&&&\\
      $\ge 0$   & $4.53 \pm 0.07 \:^{+0.35}_{-0.30} \:^{+0.58}_{-0.47} $  
                & $5.08^{+0.11}_{-0.30}$
                & $4.58 \pm 0.07 \:^{+0.38}_{-0.32} \:^{+0.61}_{-0.48}$
                & $5.27^{+ 0.11}_{- 0.32}$   \\ &&&&\\
      $\ge 1$   & $0.84 \pm 0.03 ^{+0.13}_{-0.10} \:^{+0.11}_{-0.09}$
                & $0.81^{+0.02}_{-0.04}$
                & $0.84 \pm 0.03\:^{+0.11}_{-0.09} \:^{+0.11}_{-0.09}$
                & $0.84^{+0.02}_{-0.04}$  \\&&&& \\
      $\ge 2$   & $0.21 \pm 0.01 \:^{+0.04}_{-0.03}\:^{+0.03}_{-0.02}$
                & $0.21^{+0.01}_{-0.02}$
                & $0.23 \pm 0.02\:^{+0.04}_{-0.03}\:^{+0.03}_{-0.02} $
                & $0.21^{+0.01}_{-0.02}$   \\ &&&&\\
      $\ge 3$   & $0.047 \pm 0.007 ^{+0.014}_{-0.011}\:^{+0.008}_{-0.006}$
                & $0.05 \pm 0.02$
                & $0.064 \pm 0.008\:^{+0.016}_{-0.014}\:^{+0.010}_{-0.008}$
                & $0.05 \pm 0.02$ \\ &&&&\\
      $\ge 4$   &       -  & - 
                & $0.019 \pm 0.005 \pm 0.006 \:^{+0.004}_{-0.003}$
                & - \\ &&&&\\
      
      \hline \hline
    \end{tabular}
  \caption{The measured cross section times leptonic branching ratio
    for $W$+jets in the electron and
    muon channels as a function of corrected jet multiplicity with (in order)
    statistical, systematic, and luminosity uncertainties.  The cross
    sections are quoted in a limited and well-defined
    kinematic region, described
    in the text. The measurement was not performed in the inclusive
    4-jet bin in the electron channel because of the poor
    signal-to-background ratio. Theoretical predictions from MCFM are
    also shown, with all uncertainties combined. MCFM provides NLO predictions
    for $N_{\rm{jet}} \le 2$ and a LO prediction for $N_{\rm{jet}} = 3$.}
  \label{tab:xsec_vs_multiplicity}
  \end{center}
\end{table*}

\begin{table*}[ht]
  \begin{center}
    \begin{tabular}{c|c|c|c|c}
      \hline 
      Jet               &      & MCFM  &     & MCFM \\
      multiplicity      & \Wen~& \Wen~ &\Wmn~& \Wmn~\\ 
      \hline\hline
      &&&&\\
      $\ge 1/\ge 0$ &     $0.185 \pm 0.007 ^{+0.025}_{-0.019}$
                    &     $0.159^{+0.006}_{-0.005}$
                    &     $0.183 \pm 0.007 ^{+0.023}_{-0.020}$
                    &     $0.160_{-0.005}^{+0.006}$  \\ &&&&\\
      $\ge 2/\ge 1$ &     $0.250 \pm 0.019 ^{+0.019}_{-0.010}$
                    &     $0.255^{+0.017}_{-0.022}$
                    &     $0.274 \pm 0.020 ^{+0.018}_{-0.011}$
                    &     $0.255_{-0.021}^{+0.017}$ \\ &&&&\\
      $\ge 3/\ge 2$ &     $0.224 \pm 0.037 \pm 0.022$
                    &     $0.241^{+0.108}_{-0.061}$
                    &     $0.278 \pm 0.041 ^{+0.024}_{-0.020}$
                    &     $0.242_{-0.061}^{+0.104}$ \\ &&&&\\
      $\ge 4/\ge 3$ &  -   &   -    
                    &     $0.297 \pm 0.088 ^{+0.037}_{-0.026}$ & - \\
      &&&&\\
      \hline \hline
    \end{tabular}
  \caption{The measured cross section ratio for $W$+jets in the electron and
    muon channels as a function of corrected jet multiplicity with (in order)
    statistical and systematic uncertainties.  The cross
    section ratios are quoted in a limited and well-defined
    kinematic region, described
    in the text. The measurement was not performed in the inclusive
    4-jet bin in the electron channel because of the poor
    signal-to-background ratio. Theoretical predictions from MCFM are
    also shown, with all uncertainties combined. MCFM provides NLO predictions
    for $N_{\rm{jet}} \le 2$ and a LO prediction for $N_{\rm{jet}} = 3$.}
  \label{tab:xsec_ratio}
  \end{center}
\end{table*}

Also shown in Figures \ref{fig:result-xsec},
\ref{fig:result-xsecratio}, and \ref{fig:result-xsec-jetpt} 
are particle-level expectations from PYTHIA, ALPGEN and SHERPA
simulations
as well as a calculation using MCFM v5.8 \cite{MCFM}.  PYTHIA is LO,
while ALPGEN and SHERPA match higher-multiplicity matrix elements to
a leading-logarithmic parton shower; these predictions have been
normalised to the NNLO inclusive $W$~production cross section.
The version of MCFM used here
provides NLO predictions at parton level for $W$ boson
production with $N_{\rm{jet}} \le 2$; only
leading-order predictions are available for $W$ + three jets.  No
additional normalisation was applied to the MCFM predictions.

The MCFM results were obtained with
the same jet algorithm and same kinematic selection requirements as applied
to the data.  Renormalisation and factorisation scales were set to
\HT/2, where  \HT~is the scalar sum of the \pT~of the unclustered
partons and of the lepton and neutrino from the $W$~decay.  The PDFs were 
CTEQ6L1 \cite{CTEQ6L1}  and CTEQ6.6M \cite{CTEQ6.6M}
for the LO and NLO calculations, respectively.  Corrections for
hadronisation and underlying event were computed with PYTHIA as a
function of leading and next-to-leading jet \pT.
Hadronisation and underlying event corrections ranged from $-10$\% to
$-4$\% and $+10$\% to $+4$\%, respectively,
for jet \pT~$\simeq$ 20 \GeV~to jet \pT~$>$ 80
\GeV.  The partial cancellation of hadronisation
and underlying event
corrections \cite{CSCbook} results in an overall correction of
approximately 4\%.
The effect of final state QED radiation from the electron or muon was
computed with PYTHIA and ALPGEN (both using PHOTOS) and with SHERPA,
comparing the acceptance before radiation with the acceptance after
radiation, but summing up the photons within $\Delta R=0.1$
around the lepton.
This factor ($\simeq 1-2$\%) was applied as a correction
to the MCFM prediction.

The systematic uncertainty in the MCFM cross sections
due to fragmentation was estimated by
comparing PYTHIA with HERWIG.  Underlying event uncertainties were
estimated by comparing the AMBT1 \cite{AMBT1} event generator tune
with the tune from JIMMY \cite{Jimmy} as well as by varying the
AMBT1 tune to increase the underlying event activity by
approximately 10\%.
Renormalisation and factorisation scale uncertainties
were estimated by varying the scales, in all combinations, up and
down, by factors
of two.    PDF uncertainties were computed by summing in
quadrature the  dependence on each of the 22 eigenvectors
characterising the CTEQ6.6 PDF set; the uncertanty in
$\alpha_{\rm{s}}$ was also taken into account. An alternative PDF set,
MSTW2008 \cite{MSTW2008},
with its set of 68\% C.L. eigenvectors was also examined, and the
envelope of the uncertainties from CTEQ6.6 and MSTW2008 was taken as
the final PDF uncertainty.  The total resulting uncertainties are
given in 
Tables \ref{tab:xsec_vs_multiplicity} and
\ref{tab:xsec_ratio}.

In conclusion,
this Letter presents a measurement of the $W$+jets cross section as a
function of jet multiplicity
in $pp$ collisions at $\sqrt{s} = 7$~\TeV~in
both electron and muon decay modes of the
$W$ boson, based on an integrated luminosity of 1.3 \ipb~recorded with
the ATLAS detector.
Measurements are also presented of the ratio of 
cross sections
$\sigma(W+ \ge n)/\sigma(W+ \ge n-1)$
for inclusive
jet multiplicities $n=1-4$,
and of the \pT~distribution
of the leading and next-to-leading jets in the event.
The results have been corrected for all known detector effects and are
quoted in a limited and well-defined range of jet and lepton
kinematics.  This range is fully covered by the detector acceptance,
so as to avoid model-dependent extrapolations and to facilitate
comparisons with theoretical predictions.
As expected, the PYTHIA samples considered, which contain a
$2 \rightarrow 1$ matrix element merged with a $2 \rightarrow 2$
matrix element and a leading-logarithmic parton
shower, dies not provide a good description of
the data for jet multiplicities greater
than one.
Good agreement is
observed with the predictions of the multi-parton matrix element generators
ALPGEN and SHERPA.  Calculations based on
$O(\alpha_{\rm{s}}^2)$ matrix elements in MCFM (available for jet
multiplicities $n \le 2$) are also in good agreement with the
data.

\section{Acknowledgements}

We wish to thank CERN for the efficient commissioning and operation of the LHC during this initial high-energy data-taking period as well as the support staff from our institutions without whom ATLAS could not be operated efficiently.

We acknowledge the support of ANPCyT, Argentina; YerPhI, Armenia; ARC,
Australia; BMWF, Austria; ANAS, Azerbaijan; SSTC, Belarus; CNPq and
FAPESP, Brazil; NSERC, NRC and CFI, Canada; CERN; CONICYT, Chile; CAS,
MOST and NSFC, China; COLCIENCIAS, Colombia; MSMT CR, MPO CR and VSC
CR, Czech Republic; DNRF, DNSRC and Lundbeck Foundation, Denmark;
ARTEMIS, European Union; IN2P3-CNRS, CEA-DSM / IRFU, France; GNAS,
Georgia; BMBF, DFG, HGF, MPG and AvH Foundation, Germany; GSRT,
Greece; ISF, MINERVA, GIF, DIP and Benoziyo Center, Israel; INFN,
Italy; MEXT and JSPS, Japan; CNRST, Morocco; FOM and NWO,Netherlands;
RCN,Norway;  MNiSW,Poland; GRICES and FCT, Portugal;  MERYS (MECTS),
Romania;  MES of Russia and ROSATOM, Russian Federation; JINR; MSTD,
Serbia; MSSR, Slovakia; ARRS and MVZT, Slovenia; DST / NRF, South Africa; MICINN, Spain; SRC and Wallenberg Foundation, Sweden; SER,  SNSF and Cantons of Bern and Geneva, Switzerland;  NSC, Taiwan; TAEK, Turkey; STFC, the Royal Society and Leverhulme Trust, United Kingdom; DOE and NSF, United States of America.  

The crucial computing support from all WLCG partners is acknowledged gratefully, in particular from CERN and the ATLAS Tier-1 facilities at TRIUMF (Canada), NDGF (Denmark, Norway, Sweden), CC-IN2P3 (France), KIT/GridKA (Germany), INFN-CNAF (Italy), NL-T1 (Netherlands), PIC (Spain), ASGC (Taiwan), RAL (UK) and BNL (USA) and in the Tier-2 facilities worldwide.

\bibliographystyle{model1-num-names}
\bibliography{plb}

\newpage

\begin{flushleft}
{\Large The ATLAS Collaboration}

\bigskip

G.~Aad$^{\rm 48}$,
B.~Abbott$^{\rm 111}$,
J.~Abdallah$^{\rm 11}$,
A.A.~Abdelalim$^{\rm 49}$,
A.~Abdesselam$^{\rm 118}$,
O.~Abdinov$^{\rm 10}$,
B.~Abi$^{\rm 112}$,
M.~Abolins$^{\rm 88}$,
H.~Abramowicz$^{\rm 153}$,
H.~Abreu$^{\rm 115}$
% ...
\bigskip

% ...
$^{10}$ Institute of Physics, Azerbaijan Academy of Sciences, H. Javid Avenue 33, AZ 143 Baku, Azerbaijan\\
$^{11}$ Institut de F\'isica d'Altes Energies, IFAE, Edifici Cn, Universitat Aut\`onoma  de Barcelona,  ES - 08193 Bellaterra (Barcelona), Spain\\
% ...
$^{48}$ Albert-Ludwigs-Universit\"{a}t, Fakult\"{a}t f\"{u}r Mathematik und Physik, Hermann-Herder Str. 3, D - 79104 Freiburg i.Br., Germany\\
$^{49}$ Universit\'e de Gen\`eve, Section de Physique, 24 rue Ernest Ansermet, CH - 1211 Geneve 4, Switzerland\\
% ...
$^{88}$ Michigan State University, Department of Physics and Astronomy, High Energy Physics Group, East Lansing, MI 48824-2320, United States of America\\
% ...
$^{111}$ University of Oklahoma, Homer L. Dodge Department of Physics and Astronomy, 440 West Brooks, Room 100, Norman, OK 73019-0225, United States of America\\
$^{112}$ Oklahoma State University, Department of Physics, 145 Physical Sciences Building, Stillwater, OK 74078-3072, United States of America\\
% ...
$^{115}$ LAL, Univ. Paris-Sud, IN2P3/CNRS, Orsay, France\\
% ...
$^{118}$ Oxford University, Department of Physics, Denys Wilkinson Building, Keble Road, Oxford OX1 3RH, United Kingdom\\
% ...
$^{153}$ Tel Aviv University, Raymond and Beverly Sackler School of Physics and Astronomy, Ramat Aviv, IL - Tel Aviv 69978, Israel\\
% ...
\end{flushleft}

\end{document}